\documentclass[pra,reprint,superscriptaddress,showpacs,longbibliography]{revtex4-1}
\usepackage{amssymb,amsthm,amsmath,bbm,bbold,graphicx,epsfig,threeparttable,color}%%{\color{red} text}
\usepackage[colorlinks=true,linkcolor=blue]{hyperref}
\usepackage{ulem,enumerate}
\newcommand{\ket}[1]{{| #1 \rangle}}
\newcommand{\bra}[1]{{\langle #1 |}}

\theoremstyle{definition}
\begin{document}

\title{Simulating Open Quantum Systems with Hamiltonian Ensembles and the Nonclassicality of the Dynamics}
\author{Hong-Bin Chen}
\email{hongbinchen@phys.ncku.edu.tw}
\affiliation{Department of Physics, National Cheng Kung University, Tainan 70101, Taiwan}
\author{Clemens Gneiting}
\affiliation{Quantum Condensed Matter Research Group, RIKEN, Wako-shi, Saitama 351-0198, Japan}
\author{Ping-Yuan Lo}
\affiliation{Department of Electrophysics, National Chiao Tung University, Hsinchu 30010, Taiwan}
\author{Yueh-Nan Chen}
\email{yuehnan@mail.ncku.edu.tw}
\affiliation{Department of Physics, National Cheng Kung University, Tainan 70101, Taiwan}
\affiliation{Physics Division, National Center for Theoretical Sciences, Hsinchu 30013, Taiwan}
\author{Franco Nori}
\affiliation{Quantum Condensed Matter Research Group, RIKEN, Wako-shi, Saitama 351-0198, Japan}
\affiliation{Physics Department, University of Michigan, Ann Arbor, Michigan 48109-1040, USA}

\date{\today}

\begin{abstract}
The incoherent dynamical properties of open quantum systems are generically attributed to an ongoing correlation between the system and its environment. Here, we propose a novel way to assess the nature of these system-environment correlations by examining the system dynamics alone. Our approach is based on the possibility or impossibility to simulate open-system dynamics with Hamiltonian ensembles. As we show, such (im)possibility to simulate is closely linked to the system-environment correlations. We thus define the nonclassicality of open-system dynamics in terms of the nonexistence of a Hamiltonian-ensemble simulation. This classifies any nonunital open-system dynamics as nonclassical. We give examples for open-system dynamics that are unital and classical, as well as unital and nonclassical.
\end{abstract}

\maketitle

%==============
%Introduction
%==============

\textit{Introduction}.---When a quantum system interacts with its environment, its dynamical behavior will, in general, deviate from the dynamics of a strictly isolated one
\cite{leggett_diss_sys_rmp_1987,breuer_textbook,weiss_textbook,pyl_n_marko_dyna_prl_2012,hengna_n_marko_comp_sr_2015,hongbin_scirep_2015}. As a result of an ongoing bipartite
correlation arising from the system-environment interaction, the system dynamics may display incoherent characteristics, such as dephasing or damping processes.
Formally, such processes are captured by quantum master equations, replacing the von Neumann equation for isolated systems.

\begin{figure}[ht]
\includegraphics[width=\columnwidth]{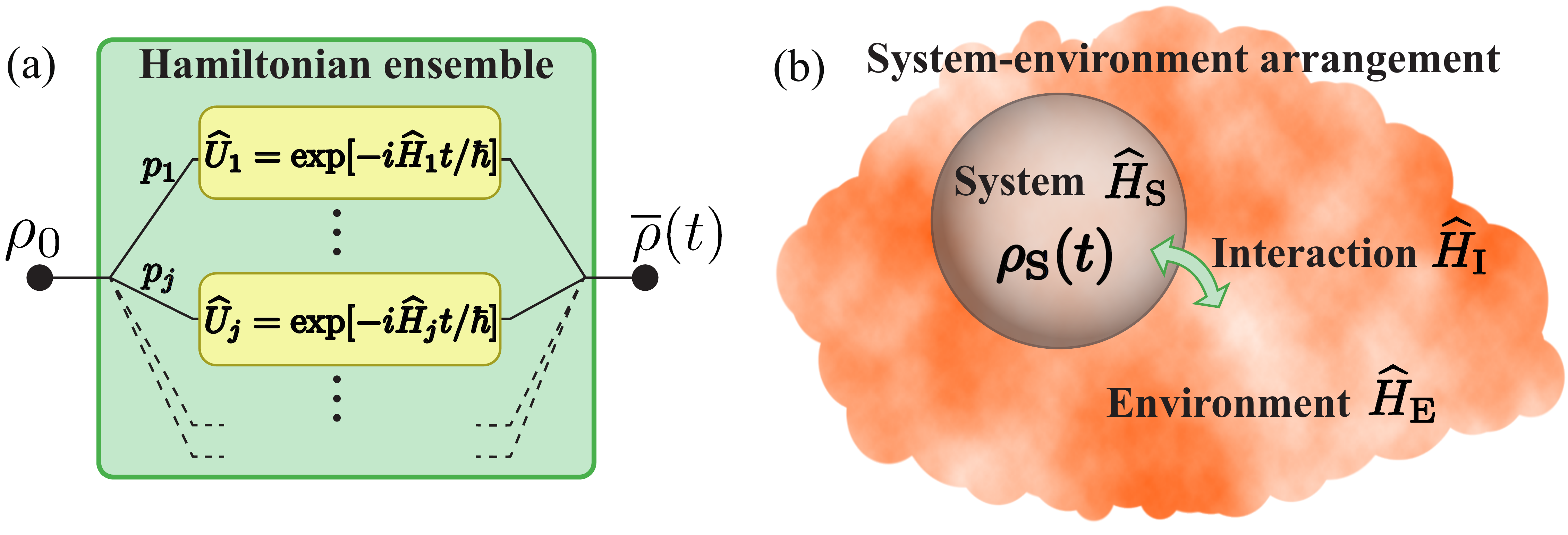}
\caption{(a) Schematic illustration showing the averaged state $\overline{\rho}(t)$ resulting from the Hamiltonian ensemble $\{(p_{j},\widehat{H}_{j})\}$. The averaged state, in general, follows incoherent dynamics; i.e., similar to open quantum systems, its time evolution cannot be accurately captured by the von Neumann equation alone. (b) If a quantum system interacts with a bath, their interaction will, in general, correlate them. As we show, the nature of these correlations, i.e., classical or quantum, is intimately connected to the (im)possibility to simulate the dynamics of the open system with a Hamiltonian ensemble.}	 \label{Fig:illustration}
\end{figure}

However, incoherent dynamics can also arise as a consequence of a purely classical averaging procedure over distinct autonomous evolutions. For example,
the double slit experiment can,
when exposed to a disordered potential and after averaging,
encounter similar decoherence as if which-slit information had leaked into an environment \cite{Gneiting2016incoherent}.
In this sense, disordered quantum systems described by Hamiltonian ensembles can behave in an analogous manner as open quantum systems,
even if individual realizations are strictly isolated (Fig.~\ref{Fig:illustration}) \cite{Gneiting2016incoherent,Kropf2016effective,clemens_disordered_pra_2017,clemens_disordered_prl_2017}.

Here, we exploit this dynamical correspondence to assess the nature of the system-environment correlations in terms of the system properties alone. As we show, the impossibility to simulate is necessarily linked to nonclassical system-environment correlations. On the other hand, if such a simulation is possible, then there always exists a system-environment model which reproduces the system dynamics by relying only on classical correlations. This leads us to defining the nonclassicality of open-system dynamics in terms of the nonexistence of a Hamiltonian-ensemble simulation.

Alternative definitions for the nonclassicality of system dynamics have been proposed \cite{rahimikeshari_process_n_cla_prl_2013,krishna_process_n_cla_pra_2016}.
In these definitions, the dynamics is considered classical if the state preserves classicality during the temporal evolution.
Typically, the classicality of states in these approaches is formulated in terms of the Wigner function or the Glauber-Sudarshan $P$ representation \cite{wigner_func_pr_1932, glauber_pr_1963, sudarshan_prl_1963, miranowicz_n_cla_pra_2010, miranowicz_n_cla_pra_2011, miranowicz_n_cla_pra_2015}. While these definitions also rely on system properties alone, their applicability is limited to systems amenable for such a phase space description, excluding other cases of interest. We, instead, propose to discuss the nonclassicality of {\it open}-system dynamics separately, based on the system-environment correlations and independent of the nature of the system.

As an immediate consequence of our definition, all nonunital dynamics, e.g., dissipative processes, are nonclassical.
The spin-boson model, in contrast, which displays unital dynamics and---on the level of the model---quantum correlations, can, as we find, be simulated and thus exhibits classical open-system dynamics according to our definition, since these correlations cannot be certified by considering the system dynamics alone. In the case of an extended spin-boson model, however, where the environment is complemented by a second qubit and the system dynamics remains unital, we prove the nonexistence of a simulating Hamiltonian ensemble for certain spectral densities; i.e., the system's evolution is in these cases witnessed to be manifestly nonclassical.

%==================================
%Dynamics of Hamiltonian ensembles
%==================================

\textit{Dynamics of Hamiltonian ensembles}.---An isolated quantum system is described by a Hamiltonian ensemble (HE) $\{(p_{j},\widehat{H}_{j})\}$, if the autonomous Hamiltonian $\widehat{H}_{j}$ of the system is drawn from a probability distribution $p_{j}$ [see Fig.~\ref{Fig:illustration}(a)].
Such HEs are applicable to describing disordered quantum systems. Here, we relate HEs to open quantum systems.

The dynamics of the ensemble averaged state $\overline{\rho}(t)$ exhibits features distinct from the dynamics of any single realization. The latter is governed by the unitary evolution
$\rho_j(t)=\widehat{U}_j\rho_0\widehat{U}_j^\dagger$, with the initial state $\rho_0$ and $\widehat{U}_j=\exp[-i\widehat{H}_jt/\hbar]$, whereas the
dynamics of the averaged state $\overline{\rho}(t)$ is given by the unital (i.e., identity invariant) map
\begin{align}
\overline{\rho}(t) = \sum_j p_j \exp[-i\widehat{H}_jt/\hbar] \rho_{0}
\exp[i\widehat{H}_jt/\hbar].
\label{Eq:Hamiltonian_ensemble}
\end{align}
Note that an evolution equation for $\overline{\rho}(t)$ cannot be reduced to some effective Hamiltonian alone but must, in general, take the form of a quantum master equation \cite{Gneiting2016incoherent, Kropf2016effective} (see \cite{exp_hami_ensem_nat_commun_2013} for an experimental implementation of this).

A seminal and instructive example considers a single qubit subject to spectral disorder; i.e., the Hamiltonians in the ensemble differ only in their eigenvalues, while they share a
common basis of eigenstates \cite{Kropf2016effective}. The HE may be given by $\{(p(\omega ),\hbar \omega \hat{\sigma}_{z}/2)\}$, with the probability
distribution $p(\omega)$ kept general. The resulting master equation reads
\begin{align}
\frac{\partial}{\partial t}\overline{\rho}(t)=-\frac{i}{\hbar}[\varepsilon(t) \hat{\sigma}_z,\overline{\rho}(t)]
+\gamma(t)[\hat{\sigma}_z\overline{\rho}(t)\hat{\sigma}_z-\overline{\rho}(t)] ,
\label{Eq:Single_qubit_spectral_disorder}
\end{align}
where the effective energy $\varepsilon(t)=\hbar\mathrm{Im}[\partial_{t}\ln\phi(t)]/2$ and the decoherence rate $\gamma(t)=-\mathrm{Re}[\partial_{t}\ln\phi(t)]/2$ follow from the
dephasing factor
\begin{equation}
\phi(t)=\int_{-\infty}^{\infty} p(\omega) e^{i\omega t}d\omega.
\label{Eq:Dephasing_factor_of_ensemble}
\end{equation}
Depending on the underlying probability distribution $p(\omega )$, the master equation~(\ref{Eq:Single_qubit_spectral_disorder}) can range from time-constant dephasing to a
strongly oscillating incoherent behavior, the latter even giving rise to purity revivals \cite{Kropf2016effective}.

It is worthwhile to recall that the occurrence of incoherent dynamics in the case of HEs is a consequence of the averaging procedure. Nevertheless, it is reminiscent of open quantum systems, where, in contrast, an ongoing correlation between the system and environment gives rise to the incoherent dynamics.
This Letter explores the possibility to simulate open quantum systems with HEs, and vice versa, and the implications on the system-environment correlations.

%==================================
%Simulating open quantum systems with Hamiltonian ensembles
%==================================
\textit{Simulating open quantum systems with Hamiltonian ensembles}.---We now show that nonclassical system-environment correlations are necessarily linked to the impossibility to simulate the
open-system dynamics with a HE. To this end, we show---conversely---that, if system and environment are persistently classically correlated, then the reduced system state is described by a HE.

A system-environment arrangement is characterized by an autonomous total Hamiltonian
$\widehat{H}_\mathrm{T}=\widehat{H}_\mathrm{S}+\widehat{H}_\mathrm{E}+\widehat{H}_\mathrm{I}$, with the system $\widehat{H}_\mathrm{S}$, the environment
$\widehat{H}_\mathrm{E}$, and the interaction Hamiltonian $\widehat{H}_\mathrm{I}$ [see Fig.~\ref{Fig:illustration}(b)]. The total system evolves unitarily as $\rho_\mathrm{T}(t) =\widehat{U}\rho_{\mathrm{T},0}\widehat{U}^\dagger$, with $\widehat{U}=\exp[-i\widehat{H}_\mathrm{T}t/\hbar]$. We say
that an open system is described by a HE if the reduced system state $\rho_\mathrm{S}(t)=\mathrm{Tr}_\mathrm{E}[\rho_\mathrm{T}(t)]$ allows a decomposition of the
form~(\ref{Eq:Hamiltonian_ensemble}), where the probabilities $p_j$ and the Hamiltonians $\widehat{H}_j$ of the ensemble are determined by $\widehat{H}_\mathrm{T}$ and the
initial state $\rho_{\mathrm{T},0}$.

Instead of  further specifying the total Hamiltonian $\widehat{H}_\mathrm{T}$, we now assume that the total state $\rho_\mathrm{T}(t)$ remains at all times classically correlated, displaying neither quantum discord \cite{quantum_discord_prl_2002,condition_nonzero_discord_prl_2010} nor entanglement.
Under this condition, we argue that, for every initial state of the form $\rho_{\mathrm{T},0}=\rho_{\mathrm{S},0}\otimes\sum_j p_j\ket{j}\bra{j}$ (with $\{p_j\}$ a time-independent probability distribution and $\{\ket{j}\}$ a basis of the environment), the reduced system state can be described by a HE.

The detailed proof is presented in the Supplemental Material \cite{supp_simu_open_sys_hamil_ensem}. Here, we outline the central steps. First, as a direct consequence of the classical correlations, there exists an environmental basis $\{\ket{k}\}$ (in general different from $\{\ket{j}\}$), such that
\begin{equation}
\rho_\mathrm{T}(t)=\sum_{k,j}p_j\widehat{E}_{k,j}\rho_{\mathrm{S},0}\widehat{E}_{k,j}^\dagger\otimes\ket{k}\bra{k},
\label{eq_rho_tot_zero_discord}
\end{equation}
where the operators $\widehat{E}_{k,j}=\bra{k}\widehat{U}(t)\ket{j}$ act on the system and satisfy $\sum_k\widehat{E}_{k,j}^\dagger\widehat{E}_{k,j}=\widehat{I}$ for each $j$.

To demonstrate that the $\widehat{E}_{k,j}$ are unitary, we again use the zero-discord assumption, which implies that each environmental off-diagonal term
vanishes, i.e.,
$\widehat{E}_{k,j}\rho_{\mathrm{S},0}\widehat{E}_{k',j}^\dagger=0$ for $k\neq k'$. This, in turn, implies that there exists a bijection between $\{\ket{j}\}$ and $\{\ket{k}\}$ such that $\widehat{E}_{k,j}$ is nonzero only when its two indices match the bijection, i.e., $\widehat{E}_{k,j}=\widehat{E}_{k_{j'},j}\delta_{j,j'}$. Unitarity of the $\widehat{E}_{k,j}$ then follows directly.

Finally, we address the time dependence of the $\widehat{E}_{k,j}$. Expressing the bijection as a unitary operator $\widehat{\mathcal{U}}(t)$ and safely neglecting the index $k$, we can recast $\widehat{U}(t)$ in a separable form:
\begin{equation}
\widehat{U}(t)=\sum_j\widehat{E}_j(t)\otimes\widehat{\mathcal{U}}(t)\ket{j}\bra{j} .
\end{equation}
The group properties of $\{\widehat{U}(t)|t\in\mathbb{R}\}$ are thus inherited by the operators $\widehat{E}_j(t)$ and $\widehat{\mathcal{U}}(t)$;
i.e., due to the time independence of the total Hamiltonian, we can write $\widehat{E}_j(t)=\exp[-i\widehat{H}_j t/\hbar]$ and $\widehat{\mathcal{U}}(t)=\sum_j\exp[-i(\theta_j t/\hbar)]\ket{j}\bra{j}$, with $\widehat{H}_j$ time-independent Hermitian
operators and $\theta_j$ real-valued constants. Consequently, Eq.~(\ref{eq_rho_tot_zero_discord}) corresponds to a time-independent HE $\{(p_j,\widehat{H}_j)\}$ when
tracing over the environment.

%==========================================================
%Simulating Hamiltonian ensembles with open quantum systems
%==========================================================

\textit{Simulating Hamiltonian ensembles with open quantum systems}.---The impossibility to simulate an open system with a HE certifies the quantum nature of the system-environment correlations. Notably, this is achieved by considering system properties alone, i.e., without explicit reference to the environment. We now show that, on the other hand, the existence of a simulating HE always admits the possibility of classical system-environment correlations; i.e., the latter cannot be excluded  by considering only the system.

We explicitly construct a system-environment arrangement which reproduces an arbitrary HE $\{(p_j,\widehat{H}_j)\}$ relying only on classical correlations. To this end, we
write $\widehat{H}_j=\widehat{\overline{H}}+\widehat{V}_j$ (with the average $\widehat{\overline{H}}=\sum_j p_j \widehat{H}_j$) and choose the interaction to be of the form $\widehat{H}_{\rm I} = \sum_j \widehat{V}_j \otimes \ket{j}\bra{j}$; i.e., it associates with each $\widehat{H}_j$ of the ensemble a distinct state $\ket{j}$ of an (arbitrary) basis of the environment. Note that the index $j$ is generic and may be continuous and/or a multi-index. The environment must then be chosen appropriately to accommodate the complexity of the HE. Moreover, we take the system Hamiltonian to be the average $\widehat{H}_{\rm S} = \widehat{\overline{H}}$ and the bath Hamiltonian to be diagonal in the same basis as $\widehat{H}_{\rm I}$, i.e., $[\widehat{H}_{\rm E},\widehat{H}_{\rm I}]=0$.

With a separable initial state,
$\rho_{\rm T}(0) = \rho_{{\rm S},0} \otimes \rho_{\rm E}$, and $\rho_{\rm E} = \sum_j p_j \ket{j} \bra{j}$ (i.e., $[\rho_{\rm E},\widehat{H}_{\rm E}]=0$, and the probabilities of the Hamiltonian ensemble are assigned to the environmental populations), the time-evolved total state reads
$\rho_{\rm T}(t)=\widehat{U}_{\rm S+I}(\rho_{\rm S,0}\otimes\rho_{\rm E})\widehat{U}^\dagger_{\rm S+I}$, with
$\widehat{U}_{\rm S+I}=\exp[-i(\widehat{H}_{\rm S}+\widehat{H}_{\rm I})t/\hbar]$. Rewriting $\widehat{U}_{\rm S+I}=\sum_j \widehat{U}_j \otimes \ket{j}\bra{j}$, with $\widehat{U}_j=\exp[-i\widehat{H}_jt/\hbar]$, we obtain
\begin{align}
\rho_{\rm T}(t) = \sum_{j} p_j e^{-i\widehat{H}_j t/\hbar} \rho_{{\rm S},0} e^{i\widehat{H}_{j} t/\hbar} \otimes
\ket{j}\bra{j}.
\label{Eq:Total_state_evolution}
\end{align}
If we now trace over the environment, $\rho_{\rm S}(t)=\sum_j\bra{j}\rho_{\rm T}(t)\ket{j}$, we recover the desired decomposition~(\ref{Eq:Hamiltonian_ensemble}) in terms of the HE $\{(p_j,\widehat{H}_j)\}$. Moreover, it
is easy to see that the total state (\ref{Eq:Total_state_evolution}) is exclusively classically correlated, as desired.

As an instructive example, we consider a pair of qubits coupled to each other via a controlled-NOT gate, where a control (C) qubit determines the operation on a target (T) qubit. If the state of the C qubit resides in the classical mixture $\rho_{\mathrm{C}}=a{| 1\rangle}{\langle 1|}+(1-a){|0\rangle}{\langle 0|}$
\cite{hongbin_k_div_diag_pra_2015}, the reduced dynamics of the T qubit will be described by the mixture of evolutions
$\rho_{\mathrm{T}}(t)=a\widehat{U}_x\rho_{\mathrm{T},0}\widehat{U}^\dagger_x+(1-a)\rho_{\mathrm{T},0}$, with $\widehat{U}_x=\exp[-iJ\hat{\sigma}_xt/2\hbar]$ and $J$ the coupling
strength. We thus recover the HE $\{(a,J\hat{\sigma}_x/2),(1-a,\widehat{I})\}$. In this example, the C qubit plays the role of environment, and the qubit pair is at most classically correlated.

%==========================================================
%Nonclassicality of the dynamics
%==========================================================

\textit{Nonclassicality of the dynamics}.---It appears natural to classify open-system dynamics according to their correlation with the environment; i.e., if the system and environment are merely classically correlated, the dynamics may be considered classical; if they are quantum correlated, one may call the dynamics nonclassical. In most cases, however, one does not have (full) access to the environment, rendering such an immediate definition problematic.

We now suggest to classify open-system dynamics by the (im)possibility to describe the system dynamics by a HE. On the one hand, this definition relies only on system properties, as desired from a practical point of view. On the other hand, as we have shown, it directly links to the system-environment correlations, as desirable from a conceptual perspective. Whenever such a simulation exists,
it is impossible to exclude classical system-environment correlations by knowledge of the system dynamics alone, and we call the latter classical. If the simulation does not exist, quantum correlations must be involved; hence, the dynamics is nonclassical.

As a direct consequence of our definition, any nonunital dynamics is classified nonclassical---a simulating HE is manifestly excluded, certifying the presence of quantum correlations. This includes, e.g., dissipative processes such as the spontaneous decay of an atom. On the other hand, according to our operational definition, we may even call an open-system dynamics classical if the actual system-environment correlations are quantum. This is because our approach is deliberately ignorant of the actual environment and relies only on the {\it possibility} to explain the system dynamics with classical correlations. Next, we give an example for this.

%==========================================================
%Simulating the spin-boson model
%==========================================================

\textit{Simulating the spin-boson model}.---We now show that the system dynamics of the spin-boson model can be simulated by a HE, even though the actual model displays quantum correlations \cite{soumya_sbm_env_ent_prb_2014,jake_rc_sbm_pra_2014}. The spin-boson model
\begin{align} \label{Eq:Spin-boson_model}
\widehat{H}_{\rm S} = \frac{\hbar \omega_0}{2} \hat{\sigma}_z \qquad &, \qquad \widehat{H}_{\rm E} = \sum_{\vec{k}} \hbar \omega_{\vec{k}} \hat{b}_{\vec{k}}^{\dagger}\hat{b}_{\vec{k}}, \nonumber \\
\widehat{H}_{\rm I} = \hat{\sigma}_z \otimes & \sum_{\vec{k}} \hbar (g_{\vec{k}} \hat{b}_{\vec{k}}^{\dagger} + g_{\vec{k}}^* \hat{b}_{\vec{k}}),
\end{align}
has been extensively studied and is analytically solvable \cite{breuer_textbook}.
Tracing over the environment, the qubit system exhibits pure dephasing dynamics characterized by the dephasing factor
\begin{equation}\label{Eq:Dephasing_factor_of_sys_env_int}
\phi(t)=\exp\left[i\omega_0 t-\Phi(t)\right].
\end{equation}
In contrast to Eq.~(\ref{Eq:Dephasing_factor_of_ensemble}), which results from averaging over a HE, the dephasing factor~(\ref{Eq:Dephasing_factor_of_sys_env_int})
incorporates the information of the interaction and the environment into
$\Phi(t)=4\int_0^\infty\omega^{-2}\mathcal{J}(\omega)\coth\left(\hbar\omega/2k_\mathrm{B}T\right)\left(1-\cos\omega t\right)d\omega$, where $\mathcal{J}(\omega)$ is the environmental
spectral density. The above solution assumes that the initial state is a direct product, and that the environment is initially thermalized at temperature $T$.

We now construct a simulating HE. In view of Eq.~(\ref{Eq:Single_qubit_spectral_disorder}), we deduce that individual member Hamiltonians in the ensemble must be of the form $\omega\hat{\sigma}_\mathrm{z}/2$, which leaves us with determining the corresponding probabilities.
Given a probability distribution $p(\omega)$, the averaged dynamics can be determined by Eq.~(\ref{Eq:Dephasing_factor_of_ensemble}). Conversely, the
underlying distribution function leading to a specific dephasing factor~(\ref{Eq:Dephasing_factor_of_sys_env_int}) is obtained via the inverse Fourier transform
\begin{equation}\label{Eq:Probability_distribution}
\wp(\omega)=\frac{1}{2\pi}\int_{-\infty}^\infty \exp[i\omega_0t-\Phi(t)]e^{-i\omega t}dt.
\end{equation}
It is clear that the effect of $\omega_0$ is merely to shift $\wp(\omega)$.

To be a legitimate probability distribution function, the resulting $\wp(\omega)$ in Eq.~(\ref{Eq:Probability_distribution})
must be normalized [$\int_{-\infty}^\infty\wp(\omega)d\omega=1$], real [$\wp(\omega)\in\mathbb{R}$], and positive [$\wp(\omega)\geq 0$].
Normalization is easily seen, since $\phi(0)=1$ follows from the fact that the pure dephasing dynamics, characterized by Eq.~(\ref{Eq:Dephasing_factor_of_sys_env_int}),
should be completely positive and trace preserving. We therefore have $\int_{-\infty}^\infty \wp(\omega)d\omega=(2\pi)^{-1}\int_{-\infty}^\infty\exp\left[i\omega_0t-\Phi(t)\right]2\pi\delta(t-0)dt=1$.
Moreover, since one is generically interested in the dynamical properties only for $t\geq0$, we can deliberately extend the time domain to the full real axis such that $\Phi(t)$ is even
and $\phi(-t)=\phi(t)^\ast$. This guarantees that $\wp(\omega)$ is real: $\wp(\omega)=(\pi)^{-1}\int_0^\infty \exp\left[-\Phi(t)\right]\cos(\omega-\omega_0)tdt \in \mathbb{R}$.

The positivity of $\wp(\omega)$ is less obvious, due to the sinusoidal factors of the integrand in Eq.~(\ref{Eq:Probability_distribution}). In the following, we invoke
\textit{Bochner's theory} \cite{Bochner_math_ann_1933} to prove the general positivity of $\wp(\omega)$. To this end, we first introduce the notion of positive definiteness. A function
$f:\mathbb{R}\rightarrow\mathbb{C}$ is called positive definite if it satisfies $\sum_{j,k}f(t_j-t_k)z_jz_k^\ast\geq0$ for any finite number of pairs
$\{(t_j,z_j)|t_j\in\mathbb{R},z_j\in\mathbb{C}\}$. Note that positive definiteness of a function is different from a positive function, since the latter may not necessarily be positive
definite and vice versa. Rather, it corresponds to the positive semidefiniteness of a Hermitian matrix $\left[f(t_j-t_k)\right]_{j,k\in\mathcal{S}}$, formed by the function values
$f(t_j-t_k)$ in accordance with a certain set of indices $\mathcal{S}$. As one can show, $\phi(t)$ in Eq.~(\ref{Eq:Dephasing_factor_of_sys_env_int}) is indeed positive definite. The
proof is given in Ref. \cite{supp_simu_open_sys_hamil_ensem}.

Bochner's theorem states that a function $f$, defined on $\mathbb{R}$, is the Fourier transform of unique positive measure with density function $\wp$, if and only if $f$ is continuous
and positive definite \cite{loomis_text_book_1953,rudin_text_book_1990}. We can thus conclude that $\phi(t)$ in Eq.~(\ref{Eq:Dephasing_factor_of_sys_env_int}) is the Fourier transform
of a certain valid probability distribution $\wp(\omega)$ [Eq.~(\ref{Eq:Probability_distribution})], i.e., an analog to Eq.~(\ref{Eq:Dephasing_factor_of_ensemble}).

In summary, we have proven that there exists a unique HE, $\left\{\left(\wp(\omega),\omega\hat{\sigma}_\mathrm{z}/2\right)\right\}$, which simulates the system dynamics exactly,
irrespective of the spectral density $\mathcal{J}(\omega)$ and the associated, possibly intricate system-environment entanglement. We thus call this dynamics classical.

%==========================================================
%Extended spin-boson model
%==========================================================

\textit{Extended spin-boson model}.---Unless the system dynamics is nonunital, proving the nonexistence of a simulating HE is, in general, a nontrivial task. We now accomplish this for an
extended spin-boson model, at the same time deducing the presence of quantum correlations from system properties alone.

Our model consists of two qubits coupled to a common boson environment. The system and the interaction Hamiltonian are replaced by
$\widehat{H}_\mathrm{S}=\sum_{j}\hbar\omega_j\hat{\sigma}_{z,j}/2$ ($j=1$,~$2$) and
$\widehat{H}_\mathrm{I}=\sum_{j,\vec{k}}\hat{\sigma}_{z,j} \otimes \hbar(g_{j,\vec{k}} \hat{b}_{\vec{k}}^{\dagger} + g_{j,\vec{k}}^* \hat{b}_{\vec{k}})$,
respectively, while the environment Hamiltonian $\widehat{H}_\mathrm{E}$ is kept as in Eq.~(\ref{Eq:Spin-boson_model}). Note that the two qubits do not interact
directly. The coupling constants $g_{j,\vec{k}}$ are, in general, complex numbers. In order to reveal the nonclassical effects caused by their relative phase, we assume, for
simplicity, that they have the same amplitude, i.e., $g_{2,\vec{k}}=g_{1,\vec{k}}e^{i\varphi}$.

In the interaction picture, the total system evolves according to
$\widehat{U}^\mathrm{I}(t)=\mathcal{T}\left\{\exp\left[-i\int_0^t\sum_{\vec{k}}\widehat{Z}_{\vec{k}}\hat{b}_{\vec{k}}^\dagger(\tau)
+\widehat{Z}_{\vec{k}}^\dagger \hat{b}_{\vec{k}}(\tau)d\tau\right]\right\}$, with $\mathcal{T}$ the time-ordering operator,
$\widehat{Z}_{\vec{k}}=\sum_{j}g_{j,\vec{k}}\hat{\sigma}_{z,j}$, and $\hat{b}_{\vec{k}}(t)=e^{-i\omega_{\vec{k}}t}\hat{b}_{\vec{k}}$, respectively.
In contrast to the conventional spin-boson model, time ordering plays a nontrivial role here \cite{reina_extended_spin_bosom_model}.
(For details, see Supplemental Material~\cite{supp_simu_open_sys_hamil_ensem}.)

In the following, we regard one qubit as the system and the other as part of the environment. The reduced dynamics of the system qubit is then pure dephasing with the dephasing
factor [cf. Eq.~(\ref{Eq:Dephasing_factor_of_sys_env_int})]
\begin{equation} \label{Eq:Extended_model_dephasing_factor}
\phi^{(\mathrm{X})}(t)=\exp\left[-i\vartheta_\varphi(t)-\Phi(t)\right],
\end{equation}
where
\begin{eqnarray}
\vartheta_\varphi(t)&=&\cos\varphi\int_0^\infty\frac{4\mathcal{J}(\omega)}{\omega^2}(\omega t-\sin\omega t)d\omega \nonumber\\
&&+\mathrm{sign}(t)\sin\varphi\int_0^\infty\frac{4\mathcal{J}(\omega)}{\omega^2}(1-\cos\omega t)d\omega.
\label{eq_vartheta}
\end{eqnarray}
In the second line, we have manually inserted $\mathrm{sign}(t)$. This ensures that $\phi^{(\mathrm{X})}(-t)=\phi^{(\mathrm{X})*}(t)$ and $\wp^{(\mathrm{X})}(\omega)\in\mathbb{R}$. The presence of $\vartheta_\varphi(t)$, however, will, in general, result in the violation of positivity. Note that, similar to the conventional spin-boson model, individual member Hamiltonians in the HE must be of the form $\omega\hat{\sigma}_\mathrm{z}/2$, which allows us to follow the same line of argument.

\begin{figure}[th]
\includegraphics[width=\columnwidth]{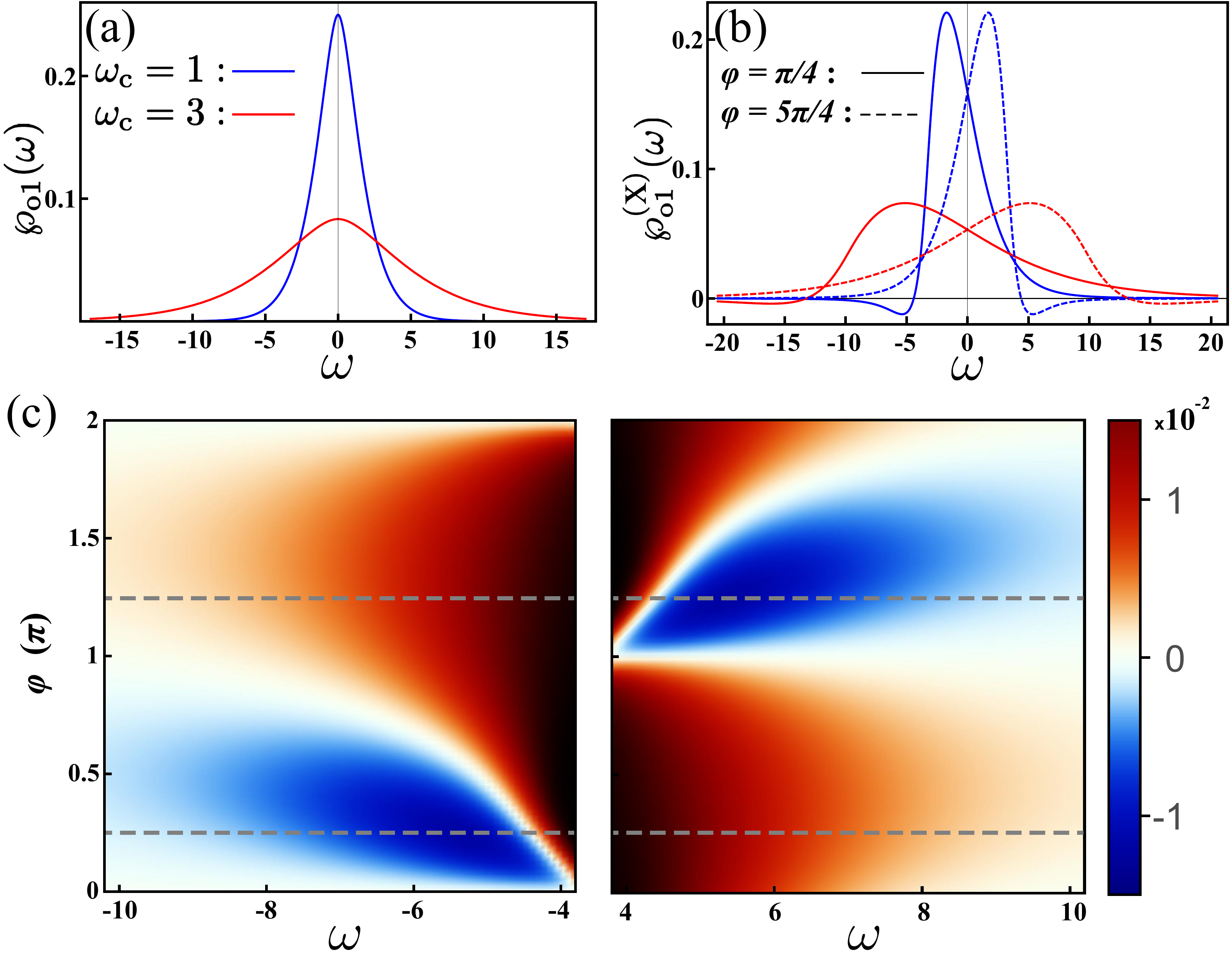}
\caption{(a) Legitimate probability distributions $\wp_\mathrm{o1}(\omega)$ for the conventional spin-boson model. (b) Distributions $\wp_\mathrm{o1}^{(\mathrm{X})}(\omega)$ for the
extended model, violating positivity. (c) The landscape of negative contributions to $\wp_\mathrm{o1}^{(\mathrm{X})}(\omega)$ against $\omega$ and $\varphi$ for $\omega_\mathrm{c}=1$.
The gray dashed lines denote the positions in Fig.~\ref{Fig:ohmic_1}(b).
}\label{Fig:ohmic_1}
\end{figure}

To demonstrate this violation, we consider the Ohmic spectral density $\mathcal{J}_\mathrm{o1}(\omega)=\omega\exp(-\omega/\omega_\mathrm{c})$ in the zero-temperature limit and a degenerate system Hamiltonian, i.e., $\omega_j=0$. In Fig.~\ref{Fig:ohmic_1}(a), we depict the legitimate probability distribution $\wp_\mathrm{o1}(\omega)$ for the conventional spin-boson model at $\omega_\mathrm{c}=1$ (blue curve) and $\omega_\mathrm{c}=3$ (red curve), while in Fig.~\ref{Fig:ohmic_1}(b), we show $\wp_\mathrm{o1}^{(\mathrm{X})}(\omega)$ for our extended model with $\varphi=\pi/4$ (solid curves) and $\varphi=5\pi/4$ (dashed curves). The latter display a manifest violation of positivity. In Fig.~\ref{Fig:ohmic_1}(c), we show the landscape of negative contributions to $\wp_\mathrm{o1}^{(\mathrm{X})}(\omega)$ against $\omega$ and $\varphi$ for $\omega_\mathrm{c}=1$. The gray dashed lines highlight $\varphi=\pi/4$ and $5\pi/4$, chosen in
Fig.~\ref{Fig:ohmic_1}(b).

%==========================================================
%Conclusions
%==========================================================

\textit{Conclusions}.---We propose a way to classify open-system dynamics according to their system-environment correlations, i.e., if the latter are classical or quantum. As we showed, this can be tested by knowledge of the system evolution alone, based on the (im)possibility to simulate the open-system dynamics with a Hamiltonian ensemble. According to our definition, any nonunital dynamics is nonclassical. Some unital system evolutions, however, such as in the spin-boson model, are classified as classical, even though the model displays quantum correlations. This highlights the operational nature of our definition.

With the extended spin-boson model, we provide an example for unital dynamics which is nonclassical according to our definition. Let us note that one may be able to simulate a larger class of
unital dynamics with time-dependent Hamiltonian ensembles. It is, for example, known that, in the case of qubits, any unital dynamics can be simulated with an ensemble of time-dependent
Hamiltonians, albeit only if also the probabilities are allowed to be time dependent \cite{landau_randon_unitary_laa_1993,audenaert_randon_unitary_njp_2008}. However, in the case of autonomous system-environment arrangements, which we consider here, such generalization appears unjustified. Finally, let us remark that demonstrating the nonexistence of a Hamiltonian-ensemble simulation is, in the case of unital evolutions, in general, a nontrivial task. An equivalent but simpler test appears desirable.

%$^{\backprime\backprime}\diagdown\textbf{\textsf{@}}$\_~\_~\_ $^{\backprime\backprime}\diagdown\textbf{\textsf{@}}$\_~\_~\_

%==========================================================
%Acknowledgements
%==========================================================

\textit{Acknowledgements}.---This work is supported partially by the National Center for Theoretical Sciences and Ministry of Science and Technology, Taiwan, Grants No. MOST
103-2112-M-006-017-MY4 and No. MOST 105-2811-M-006-059, the MURI Center for Dynamic Magneto-Optics via AFOSR Award No. FA9550-14-1-0040, the Japan
Society for the Promotion of Science (KAKENHI), the IMPACT program of JST, JSPS-RFBR Grant No. 17-52-50023, CREST Grant No. JPMJCR1676, RIKEN-AIST Challenge Research Fund, and the Sir John Templeton Foundation.

%\nocite{apsrev41Control}
%\bibliographystyle{apsrev4-1}
%\bibliography{simu_open_sys_hamil_ensem}

%

\newpage
\onecolumngrid
\clearpage
\setcounter{equation}{0}
\setcounter{figure}{0}
\begin{center}
{\bf \large Supplemental material: Simulating Open Quantum Systems with Hamiltonian Ensembles and the Nonclassicality of the Dynamics}
\end{center}
\twocolumngrid

\section{TIME-INDEPENDENT HAMILTONIAN ENSEMBLE}

In the following, we elaborate in detail the proof that classical bipartite correlations allow for a time-independent Hamiltonian ensemble decomposition of the reduced system dynamics. In this proof, we do not assume a specific form of the total Hamiltonian $\widehat{H}_\mathrm{T}$. To be precise, we now show that, if the total Hamiltonian $\widehat{H}_\mathrm{T}$ is time-independent, and if, for every initial state of the form
$\rho_{\mathrm{T},0}=\rho_{\mathrm{S},0}\otimes\sum_j p_j\ket{j}\bra{j}$ (with $\{p_j\}$ being any time-independent probability distribution), the time-evolved total
state $\rho_\mathrm{T}(t)=\widehat{U}(t)\rho_{\mathrm{T},0}\widehat{U}^\dagger(t)$ is always classically correlated between system and environment, displaying
neither quantum discord nor entanglement at any time, then the reduced system dynamics admits a time-independent Hamiltonian ensemble decomposition.

\begin{proof}
Due to the zero-discord assumption, there exists an environmental basis $\{\ket{k}\}$ (in general time-dependent and different from $\{\ket{j}\}$), such that
\begin{equation}
\rho_\mathrm{T}(t)=\sum_{k,j}p_j\widehat{E}_{k,j}\rho_{\mathrm{S},0}\widehat{E}_{k,j}^\dagger\otimes\ket{k}\bra{k},
\label{eq_rho_tot_zero_discord}
\end{equation}
where $\widehat{E}_{k,j}=\bra{k}\widehat{U}(t)\ket{j}$ are operators acting on the system Hilbert space satisfying
$\sum_k\widehat{E}_{k,j}^\dagger\widehat{E}_{k,j}=\widehat{I}$ for each $j$. At this point, we are not yet clear about the time-dependence of $\widehat{E}_{k,j}$ and
$\ket{k}$ nor the unitarity of $\widehat{E}_{k,j}$.

Crucially, the condition
\begin{equation}
\sum_j p_j\widehat{E}_{k,j}\rho_{\mathrm{S},0}\widehat{E}_{k',j}^\dagger=0
\label{eq_vanish_comp_in_rho_tot}
\end{equation}
should hold for any $k\neq k'$, due to the zero-discord assumption. Therefore each term $\widehat{E}_{k,j}\rho_{\mathrm{S},0}\widehat{E}_{k',j}^\dagger$ in the above
equation vanishes individually. The only possibility to reconcile Eqs.~(\ref{eq_rho_tot_zero_discord}) and (\ref{eq_vanish_comp_in_rho_tot}) is the existence of a
specific bijection between $\{\ket{j}\}$ and $\{\ket{k}\}$, such that $\widehat{E}_{k,j}=\widehat{E}_{k_{j'},j}\delta_{j,j'}$ for each $j$, i.e., $\widehat{E}_{k,j}$
is non-zero only when its two indices match the bijection. Then the unitarity of $\widehat{E}_{k_j,j}$ can then be confirmed according to
\begin{equation}
\sum_k\widehat{E}_{k,j}^\dagger\widehat{E}_{k,j}=\widehat{E}_{k_j,j}^\dagger\widehat{E}_{k_j,j}=\widehat{I},~\forall~j.
\label{eq_unitarity_E}
\end{equation}

The bijection between $\{|j\rangle\}$ and $\{|k\rangle\}$ can be expressed in terms of a unitary operator $\widehat{\mathcal{U}}(t)$, such that
$\bra{k_{j'}}\widehat{\mathcal{U}}(t)\ket{j}=\delta_{j,j'}$. The unitary evolution operator can then be recast in a separable form,
\begin{equation}
\widehat{U}(t)=\sum_j\widehat{E}_j(t)\otimes\widehat{\mathcal{U}}(t)\ket{j}\bra{j}.
\label{eq_u_separable_form}
\end{equation}
In the following discussion, we can, in order to keep the notation simple, safely neglect the index $k$.

Since $\{\widehat{U}(t)=\exp[-i\widehat{H}_\mathrm{T}t/\hbar]|t\in\mathbb{R}\}$ forms a group isomorphism on $\mathbb{R}$, we have the one-parameter group property
\begin{equation}
\widehat{U}(t+\delta t)=\widehat{U}(t)\widehat{U}(\delta t)
\label{eq_u_group}
\end{equation}
for $t\in\mathbb{R}$ and infinitesimal $\delta t$. Due to the unitarity of $\widehat{\mathcal{U}}(t)$, it can be expressed in terms of an Hermitian generator
$\widehat{L}(t)$ in the $\mathfrak{u}(\mathrm{dim}\mathcal{H}_\mathrm{E})$ Lie algebra on the environmental Hilbert space $\mathcal{H}_\mathrm{E}$ such that
$\widehat{\mathcal{U}}(t)=\exp[-i\widehat{L}(t)/\hbar]$. Together with
Eq.~(\ref{eq_u_separable_form}), the left hand side of Eq.~(\ref{eq_u_group}) can be written as
\begin{eqnarray}\label{eq_u_t+delta_t}
\widehat{U}(t+\delta t)=\sum_j&&\widehat{E}_j(t+\delta t) \\
&&\otimes\left[\widehat{\mathcal{U}}(t)+\frac{\partial \widehat{\mathcal{U}}(t)}{\partial t}\delta t+\mathcal{O}(\delta t^2)\right]\ket{j}\bra{j}. \nonumber
\end{eqnarray}
On the right hand side of Eq.~(\ref{eq_u_t+delta_t}), we expand $\widehat{\mathcal{U}}(t+\delta t)$ around $t$ to first order in $\delta t$. Notably, since we do not know the time-dependence and commutativity of $\widehat{L}(t)$ at this point, we can only achieve a formal expansion in Eq.~(\ref{eq_u_t+delta_t}).

Meanwhile, the right hand side of Eq.~(\ref{eq_u_group}) reads
\begin{eqnarray}\label{eq_u_t_u_delta_t}
\widehat{U}(t)\widehat{U}(\delta t)=\sum_{j',j}&&\widehat{E}_{j'}(t)\widehat{E}_j(\delta t)
\otimes\widehat{\mathcal{U}}(t)\bigg[\ket{j'}\bra{j}\delta_{j',j} \\
&&-\frac{i}{\hbar}\ket{j'}\bra{j'}\frac{\partial\widehat{L}(0)}{\partial t}\ket{j}\bra{j}\delta t+\mathcal{O}(\delta t^2)\bigg]. \nonumber
\end{eqnarray}
We again expand $\widehat{\mathcal{U}}(\delta t)$ around $t=0$. However, unlike the formal expansion in Eq.~(\ref{eq_u_t+delta_t}), we now obtain an explicit expansion in Eq.~(\ref{eq_u_t_u_delta_t}), since $\widehat{\mathcal{U}}(0)=\widehat{I}$ commutes with any operator.

Comparing Eqs.~(\ref{eq_u_t+delta_t}) and (\ref{eq_u_t_u_delta_t}), we conclude from their first terms that the group property
$\widehat{E}_j(t+\delta t)=\widehat{E}_j(t)\widehat{E}_j(\delta t)$ holds and, combined with the unitarity inferred in Eq.~(\ref{eq_unitarity_E}), that time-independent Hermitian operators $\widehat{H}_j$ exist, such that $\widehat{E}_j(t)=\exp[-i\widehat{H}_j t/\hbar]$, as well.

To reconcile the second terms of Eqs.~(\ref{eq_u_t+delta_t}) and (\ref{eq_u_t_u_delta_t}), $\partial\widehat{L}(0)/\partial t$ should be diagonalized in the basis $\{\ket{j}\}$, such that $\partial\widehat{L}(0)/\partial t=\sum_j(\partial\theta_j(0)/\partial t)\ket{j}\bra{j}$, with real parameters $\theta_j(t)$. Moreover, $\widehat{\mathcal{U}}(t)$ should satisfy
\begin{equation}
\frac{\partial\widehat{\mathcal{U}}(t)}{\partial t}=\widehat{\mathcal{U}}(t)\left[-\frac{i}{\hbar}\frac{\partial\widehat{L}(0)}{\partial t}\right].
\end{equation}
To guarantee its validity, $\partial\widehat{L}(0)/\partial t$ should commute with $\widehat{L}(t)$, since the latter is the generator of $\widehat{\mathcal{U}}(t)$. Consequently, the time-dependence of each $\theta_j(t)$ can be of first order, such that $\widehat{\mathcal{U}}(t)=\sum_j\exp[-i(\theta_j t/\hbar)]\ket{j}\bra{j}$,
with real constants $\theta_j$.
\end{proof}

Consequently, the total state in Eq.~(\ref{eq_rho_tot_zero_discord}) can be rewritten as
\begin{equation}
\rho_\mathrm{T}(t)=\sum_j p_j\widehat{U}_j\rho_{\mathrm{S},0}\widehat{U}_j^\dagger\otimes\ket{j}\bra{j},
\end{equation}
with $\widehat{U}_j=\exp[-i\widehat{H}_jt/\hbar]$, which corresponds to a time-independent Hamiltonian ensemble $\{(p_j,\widehat{H}_j)\}$ when tracing over the environment.

Finally, let us remark that, while we restrict ourselves to a time-independent total Hamiltonian, some of our conclusions
can be easily generalized to the time-dependent case. This is because Eqs.~(\ref{eq_rho_tot_zero_discord}-\ref{eq_u_separable_form}) are consequences of the zero-discord
assumption alone, regardless of the time-dependence of the total Hamiltonian. Therefore, we can also achieve the ensemble form with time-varying member Hamiltonians for a time-dependent
total Hamiltonian. However, as discussed in the main article, in the case of autonomous system-environment arrangements, i.e., in the absence of external control, such generalization
appears unjustified.

Additionally, we note that, for the case of time-independent total Hamiltonians, the separable form~(\ref{eq_u_separable_form}) not only guarantees a persistently classically correlated total
state, but also keeps the environmental basis intact without rotation, up to a phase angle $\theta_j t$.

\section{POSITIVE DEFINITENESS}

Here, we present the proof of the positive definiteness of the dephasing factor $\phi(t)=\exp\left[i\omega_0 t-\Phi(t)\right]$. For completeness, we recall the definition of positive definiteness.

{\it Positive definiteness:}
A function $f$ defined on $\mathbb{R}$ is called positive definite if it satisfies
\begin{equation}
\sum_{j,k}f(t_j-t_k)z_jz_k^\ast\geq0
\label{eq_criterion_for_pd}
\end{equation}
for any finite number of pairs $\{(t_j,z_j)|t_j\in\mathbb{R},z_j\in\mathbb{C}\}$.

We now show that, if $\phi(t)=\exp\left[i\omega_0 t-\Phi(t)\right]$ defines a CPTP pure dephasing dynamics, and if $\Phi(t)$ is even and $\phi(-t)=\phi(t)^\ast$, then $\phi(t)$ defined on $\mathbb{R}$ is positive definite.

Note that $\phi(t)$ describing a CPTP pure dephasing dynamics implies that $\phi(0)=1$, $\Phi(0)=0$, and $|\phi(t)|\leq\phi(0)$ for any $t>0$. This means that the coherence of the system can never exceed its initial value.
These properties will be frequently used in the following proof.

\begin{proof}
To simplify the problem, we first observe that the positive definiteness of $\phi(t)$ is equivalent to that of $\exp\left[-\Phi(t)\right]$, since
\begin{eqnarray}
&&\sum_{j,k}\phi(t_j-t_k)z_j z_k^\ast= \nonumber\\
&&\sum_{j,k}\exp\left[-\Phi(t_j-t_k)\right]\left(e^{i\omega_0 t_j}z_j\right)\left(e^{i\omega_0 t_k}z_k\right)^\ast.
\label{eq_real_deph_fac}
\end{eqnarray}
Correspondingly, we can assume that $\omega_0=0$ without loss of generality.

Since Eq.~(\ref{eq_criterion_for_pd}) must be valid for any number of pairs, we give the proof in an inductive manner.

In the case of only one pair $(t_1,z_1)$, Eq.~(\ref{eq_criterion_for_pd}) is trivially satisfied. We therefore start with the case of two pairs.
As stated in the main article, Eq.~(\ref{eq_criterion_for_pd}) is equivalent to the positive semidefiniteness of the Hermitian matrix:
\begin{equation}
\mathcal{M}^{(2)}=
\left[\begin{array}{cc}
1 & \exp\left[-\Phi(t_2-t_1)\right] \\
\exp\left[-\Phi(t_1-t_2)\right] & 1
\end{array}\right].
\end{equation}
It is automatically satisfied according to the CPTP dynamics defined by $\phi(t)$.

\begin{figure}[th]
\includegraphics[width=0.5\columnwidth]{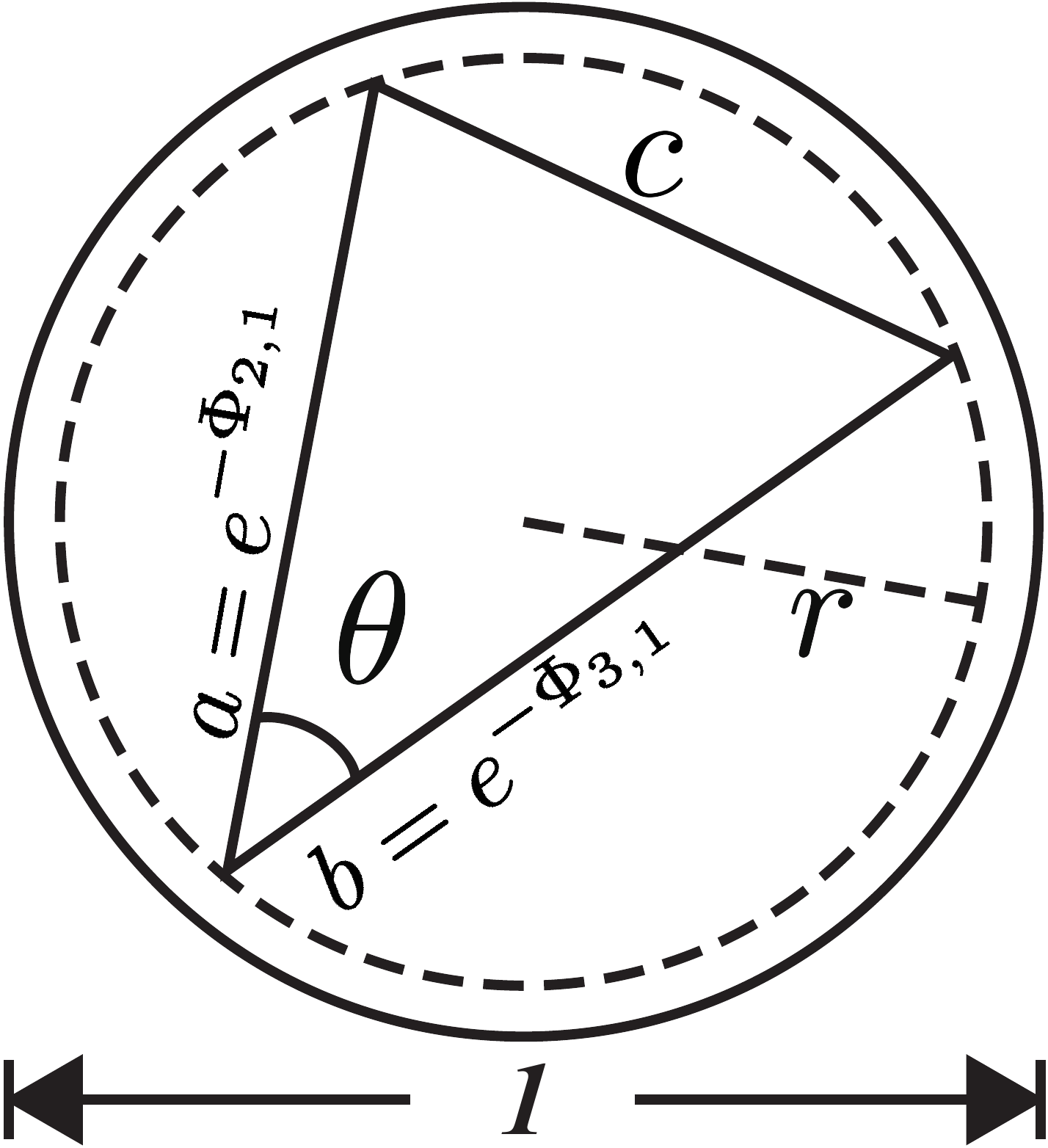}
\caption{A geometric visualization of Eq.~(\ref{eq_det_M_3}). $a$ and $b$ can be considered as two sides of a triangle with angle $\theta$ and
circumcircle (dashed circle) of diameter $2r$ less than $1$. They are all enclosed in the circle (solid circle) of diameter $1$.}
\label{fig_geo_visu_law_sines}
\end{figure}

\begin{figure*}[th]
\includegraphics[width=0.7\textwidth]{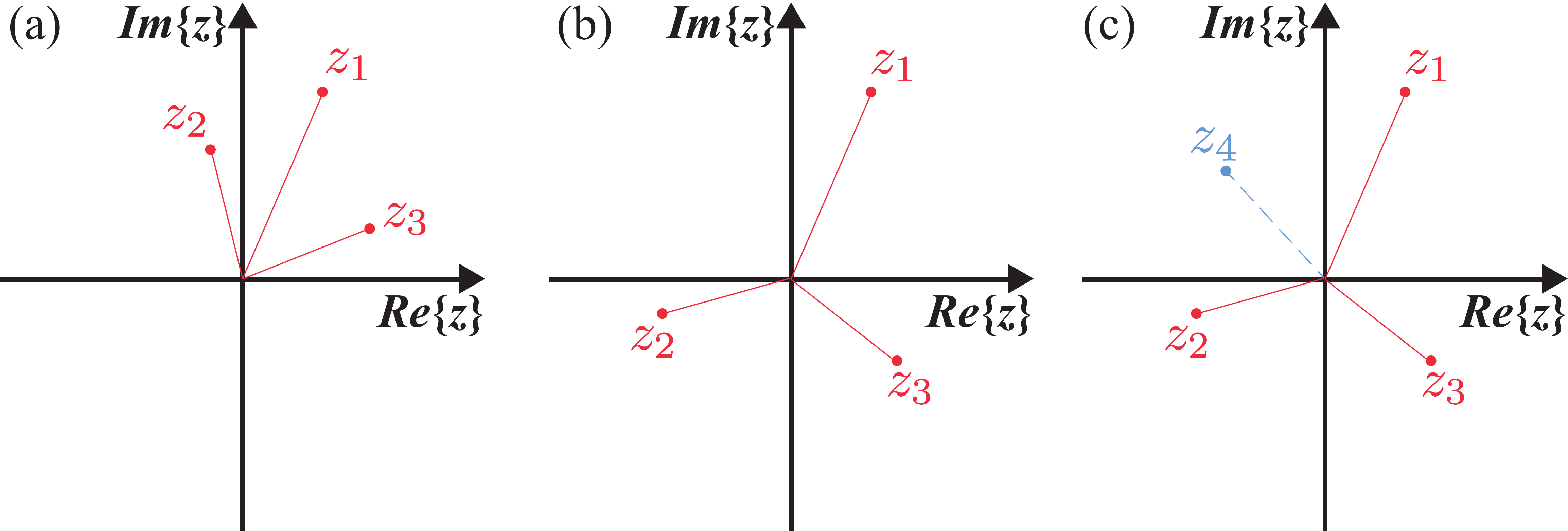}
\caption{(a) If the angles between any two $z_j$ is less than $\pi/2$, the summation of all off-diagonal elements in array~(\ref{eq_array_3_pairs})
is positive. (b) To maximize the negative contributions of off-diagonal elements, we must choose appropriate $z_j$, such that all the relative
arguments strictly exceed $\pi/2$. (c) In the case of four pairs, it is impossible to insert the fourth $z_4$ such that all relative arguments
are strictly larger than $\pi/2$.}
\label{fig_geo_visu_complex_plane}
\end{figure*}

We proceed to show the positive semidefiniteness of the Hermitian matrix
\begin{equation}
\mathcal{M}^{(3)}=
\left[\begin{array}{ccc}
1 & e^{-\Phi_{2,1}} & e^{-\Phi_{3,1}} \\
e^{-\Phi_{1,2}} & 1 & e^{-\Phi_{3,2}} \\
e^{-\Phi_{1,3}} & e^{-\Phi_{2,3}} & 1
\end{array}\right],
\end{equation}
for the case of three pairs. In the above matrix, and hereafter, the abbreviation $\Phi_{j,k}=\Phi(t_j-t_k)$ has been adopted.
Since $\mathcal{M}^{(3)}$ is three-dimensional, it is generically hard to write down an analytic expression for its three
eigenvalues $\lambda_\mu$. Nevertheless, analyzing its characteristic polynomial gives us substantial knowledge on the eigenvalues:

\begin{enumerate}[(i)]
\item $\lambda_1+\lambda_2+\lambda_3=3\geq0$ follows from the invariance of the trace.
\item $\lambda_1\lambda_2+\lambda_2\lambda_3+\lambda_3\lambda_1$ equals to the sum of all principal minors of $\mathcal{M}^{(3)}$ of order $2$ and is
consequently non-negative, since each principal minor is non-negative, following the positive semidefiniteness of $\mathcal{M}^{(2)}$.
\item $\lambda_1\lambda_2\lambda_3=\mathrm{det}\left(\mathcal{M}^{(3)}\right)$. The positivity of the product of eigenvalues is verified with the help of a simple geometric visualization shown in Fig.~\ref{fig_geo_visu_law_sines}. Explicitly expanding the determinant leads to
\begin{eqnarray}
\mathrm{det}\left(\mathcal{M}^{(3)}\right)&=&\left(1-\cos^2\theta\right)-\left(a^2+b^2-2ab\cos\theta\right) \nonumber\\
&=&\sin^2\theta-c^2, \label{eq_det_M_3}
\end{eqnarray}
with the notation $\cos\theta=\exp\left[-\Phi_{3,2}\right]$, $a=\exp\left[-\Phi_{2,1}\right]$, and $b=\exp\left[-\Phi_{3,1}\right]$.
This can be interpreted in terms of a triangle with circumcircle (dashed circle) of diameter $2r$ less than $1$. With the help of $c/\sin\theta=2r$, the positivity of Eq.~(\ref{eq_det_M_3}) and, consequently, of the product of eigenvalues is then inferred.
\end{enumerate}

Combining (i)-(iii), we can conclude that the three eigenvalues are non-negative each and, therefore, that $\mathcal{M}^{(3)}$ is positive semidefinite.

Before proceeding to the case of four pairs, it is worthwhile to discuss how the minimum of Eq.~(\ref{eq_criterion_for_pd}) is achieved.
For the case of three pairs, the LHS of Eq.~(\ref{eq_criterion_for_pd}) is equivalent to the summation over entries in the following array:
\begin{equation}
\begin{array}{ccc}
|z_1|^2 & e^{-\Phi_{2,1}}z_2z_1^\ast & e^{-\Phi_{3,1}}z_3z_1^\ast \\
e^{-\Phi_{1,2}}z_1z_2^\ast & |z_2|^2 & e^{-\Phi_{3,2}}z_3z_2^\ast \\
e^{-\Phi_{1,3}}z_1z_3^\ast & e^{-\Phi_{2,3}}z_2z_3^\ast & |z_3|^2
\end{array}.
\label{eq_array_3_pairs}
\end{equation}
If we first determine the amplitudes $|z_j|$ and adjust their arguments and $t_j$, it is clear that the diagonal elements in the array~(\ref{eq_array_3_pairs})
are all positive and, to reduce the resulting summation, the possible negative contributions are given by the off-diagonal elements.
If we choose three pairs such that the angles between any two $z_j$ within them is less than $\pi/2$, as show in Fig.~\ref{fig_geo_visu_complex_plane}(a),
the summation of all off-diagonal elements is positive. Therefore, we must choose appropriate pairs such that all their relative arguments strictly exceed
$\pi/2$, as show in Fig.~\ref{fig_geo_visu_complex_plane}(b). To maximize the negative contributions, we assume $t_1=t_2=t_3$ and $\exp\left[-\Phi_{j,k}\right]=1$.
We therefore draw the conclusion that $\sum_{j,k}f(t_j-t_k)z_jz_k^\ast\geq|z_1+z_2+z_3|^2$ for the case of maximized relative arguments between three $z_j$.

However, in the case of four pairs, it is impossible to insert the fourth $z_4$ such that all relative arguments are strictly larger than $\pi/2$,
as shown in Fig.~\ref{fig_geo_visu_complex_plane}(c). According to the above discussion, to deal with the array of four pairs,
\begin{equation}
\begin{array}{cccc}
|z_1|^2 & e^{-\Phi_{2,1}}z_2z_1^\ast & e^{-\Phi_{3,1}}z_3z_1^\ast & e^{-\Phi_{4,1}}z_4z_1^\ast \\
e^{-\Phi_{1,2}}z_1z_2^\ast & |z_2|^2 & e^{-\Phi_{3,2}}z_3z_2^\ast & e^{-\Phi_{4,2}}z_4z_2^\ast \\
e^{-\Phi_{1,3}}z_1z_3^\ast & e^{-\Phi_{2,3}}z_2z_3^\ast & |z_3|^2 & e^{-\Phi_{4,3}}z_4z_3^\ast \\
e^{-\Phi_{1,4}}z_1z_4^\ast & e^{-\Phi_{2,4}}z_2z_4^\ast & e^{-\Phi_{3,4}}z_3z_4^\ast & |z_4|^2
\end{array},
\label{eq_array_4_pairs}
\end{equation}
we can at most group three $z_j$ with all three relative arguments strictly larger than $\pi/2$ by setting their corresponding $t_j$ equal.
Then the array~(\ref{eq_array_4_pairs}) reduces to a simpler one:
\begin{equation}
\begin{array}{cc}
|z_1+z_2+z_3|^2 & e^{-\Phi_{4,1}}z_4(z_1+z_2+z_3)^\ast \\
e^{-\Phi_{1,4}}(z_1+z_2+z_3)z_4^\ast & |z_4|^2
\end{array}.
\end{equation}
Again, in accordance with the positive semidefiniteness of $\mathcal{M}^{(2)}$, we can guarantee the validity of Eq.~(\ref{eq_criterion_for_pd})
for the case of four pairs.

For the case of five or more pairs, a similar procedure can be applied to continuously reduce the problem to an equivalent $\mathcal{M}^{(2)}$ or
$\mathcal{M}^{(3)}$ case. This implies the validity of Eq.~(\ref{eq_criterion_for_pd}) for the general case.
\end{proof}

Let us remark that the above proof
already indicates the general impossibility
of a Hamiltonian ensemble description for arbitrary pure dephasing dynamics.
Many conclusions in the above proof hold since the phase angle of $\phi(t)$ is directly proportional to time $t$. This is particularly manifest in
Eq.~(\ref{eq_real_deph_fac}). However, this is in general not the case, e.g., in the extended spin-boson model below. We consequently may obtain invalid (or quasi-) distributions in the extended spin-boson model.

\section{EXTENDED SPIN-BOSON MODEL}

We proceed with the details of the extended spin-boson model, which consists of two qubits coupled to a common boson environment. The system and the
interaction Hamiltonian of the conventional spin-boson model are thus replaced by
\begin{eqnarray}
\widehat{H}_\mathrm{S}&=&\sum_{j=1,2}\frac{\hbar\omega_j}{2}\hat{\sigma}_{z,j}, \nonumber\\
\widehat{H}_\mathrm{I}&=&\sum_{j,\vec{k}}\hat{\sigma}_{z,j} \otimes \hbar(g_{j,\vec{k}} \hat{b}_{\vec{k}}^{\dagger} + g_{j,\vec{k}}^* \hat{b}_{\vec{k}}).
\end{eqnarray}
Note that the two qubits do not interact with each other directly. Let us remark that, while we consider two qubits here, our treatment can straightforwardly be generalized to more than two qubits.

Transforming to the interaction picture with respect to $\widehat{H}_\mathrm{S}+\widehat{H}_\mathrm{E}$, the total system evolves according to the unitary evolution operator
\begin{equation}
\widehat{U}^\mathrm{I}(t)=\mathcal{T}\left\{\exp\left[-i\int_0^t\sum_{\vec{k}}\widehat{Z}_{\vec{k}}\hat{b}_{\vec{k}}^\dagger(\tau)
+\widehat{Z}_{\vec{k}}^\dagger \hat{b}_{\vec{k}}(\tau)d\tau\right]\right\},
\end{equation}
where $\mathcal{T}$ is the time-ordering operator, $\widehat{Z}_{\vec{k}}=\sum_{j=1,2}g_{j,\vec{k}}\hat{\sigma}_{z,j}$, and
$\hat{b}_{\vec{k}}(t)=e^{-i\omega_{\vec{k}}t}\hat{b}_{\vec{k}}$, respectively. In the conventional spin-boson model with a single qubit, time-ordering $\mathcal{T}$ plays no significant role, since it merely introduces a global phase to the unitary evolution operator. However, this is not the case for extended models with more than one qubit, where one must carefully deal with the effect of time-ordering $\mathcal{T}$. We therefore have
\begin{equation}
\widehat{U}^\mathrm{I}(t)=\exp\left[-i\int_0^t\sum_{\vec{k}}\widehat{Z}_{\vec{k}}e^{i\omega_{\vec{k}}\tau}\hat{b}_{\vec{k}}^\dagger d\tau\right]
\times\widehat{\mathcal{A}}(t),
\end{equation}
with
\begin{eqnarray}
\widehat{\mathcal{A}}(t)&=&\mathcal{T}\left\{\exp\left[-i\int_0^t d\tau \left(e^{i\int_0^\tau\sum_{\vec{k}}\widehat{Z}_{\vec{k}}\hat{b}_{\vec{k}}^\dagger(s)ds}\right)\right.\right. \\
&&\times\left.\left.\sum_{\vec{k}}\widehat{Z}_{\vec{k}}^\dagger e^{-i\omega_{\vec{k}}\tau}\hat{b}_{\vec{k}}
\left(e^{-i\int_0^\tau\sum_{\vec{k}}\widehat{Z}_{\vec{k}}\hat{b}_{\vec{k}}^\dagger(s)ds}\right)\right]\right\}. \nonumber
\end{eqnarray}
By using the prescription $e^{\beta\hat{b}^\dagger}\hat{b}e^{-\beta\hat{b}^\dagger}=\hat{b}-\beta$, the operator $\widehat{\mathcal{A}}(t)$
can be recast into
\begin{eqnarray}
\widehat{\mathcal{A}}(t)&=&\exp\left[-i\int_0^t d\tau \sum_{\vec{k}}\widehat{Z}_{\vec{k}}^\dagger e^{-i\omega_{\vec{k}}\tau}\right. \nonumber\\
&&\qquad\qquad\qquad\times\left.\left(\hat{b}_{\vec{k}}-i\int_0^\tau\widehat{Z}_{\vec{k}}e^{i\omega_{\vec{k}}s}ds\right)\right] \nonumber\\
&=&\exp\left[-i\int_0^t\sum_{\vec{k}}\widehat{Z}_{\vec{k}}^\dagger e^{-i\omega_{\vec{k}}\tau}\hat{b}_{\vec{k}}d\tau\right]\times
\widehat{\mathcal{B}}(t).
\end{eqnarray}
with
\begin{equation}
\widehat{\mathcal{B}}(t)=\exp\left[-\int_0^t\int_0^\tau\sum_{\vec{k}}\widehat{Z}_{\vec{k}}\widehat{Z}_{\vec{k}}^\dagger e^{-i\omega_{\vec{k}}(\tau-s)}dsd\tau\right].
\end{equation}
Given that both $\widehat{A}$ and $\widehat{B}$ commute with $\left[\widehat{A},\widehat{B}\right]$, they satisfy
$e^{\widehat{A}}e^{\widehat{B}}=e^{\left[\widehat{A},\widehat{B}\right]/2}e^{\widehat{A}+\widehat{B}}$. Then $\widehat{U}^\mathrm{I}(t)$
can easily be calculated:
\begin{eqnarray}
\widehat{U}^\mathrm{I}(t)&=&\exp\left[\frac{1}{2}\int_0^t\int_0^t\sum_{\vec{k}}\widehat{Z}_{\vec{k}}\widehat{Z}_{\vec{k}}^\dagger e^{i\omega_{\vec{k}}(\tau-s)}dsd\tau\right] \nonumber\\
&&\times\exp\left[-i\int_0^t\sum_{\vec{k}}\widehat{Z}_{\vec{k}}\hat{b}_{\vec{k}}^\dagger(\tau)+\widehat{Z}_{\vec{k}}^\dagger\hat{b}_{\vec{k}}(\tau)d\tau\right]
\times\widehat{\mathcal{B}}(t) \nonumber\\
&=&\exp\left[i\sum_{\vec{k}}\widehat{Z}_{\vec{k}}\widehat{Z}_{\vec{k}}^\dagger\left(\frac{\omega_{\vec{k}}t-\sin\omega_{\vec{k}}t}{\omega_{\vec{k}}^2}\right)\right] \nonumber\\
&&\times\exp\left[\sum_{\vec{k}}\widehat{Z}_{\vec{k}}\alpha_{\vec{k}}(t)\hat{b}_{\vec{k}}^\dagger-\widehat{Z}_{\vec{k}}^\dagger\alpha_{\vec{k}}^*(t)\hat{b}_{\vec{k}}\right],
\end{eqnarray}
where $\alpha_{\vec{k}}(t)=-i\int_0^t e^{i\omega_{\vec{k}}\tau}d\tau=\left(1-e^{i\omega_{\vec{k}}t}\right)/\omega_{\vec{k}}$.

Assuming the direct-product initial state
\begin{equation}
\rho_\mathrm{T}(0)=\rho_1(0)\otimes\rho_2(0)\otimes\rho_\mathrm{E}(0),
\end{equation}
the reduced dynamics of qubit-1, which we now consider to be our system, can be obtained by
\begin{equation}
\rho_1^\mathrm{I}(t)=\mathrm{Tr}_{2,\mathrm{E}}\left[\widehat{U}^\mathrm{I}(t)\rho_\mathrm{T}(0)\widehat{U}^{\mathrm{I}\dagger}(t)\right].
\end{equation}
The superscript $\mathrm{I}$ reminds that the dynamics is formulated in the interaction picture. One can easily show that the reduced dynamics of each qubit describes pure dephasing. We can thus apply the same method for constructing a Hamiltonian ensemble as for the conventional spin-boson model. We therefore focus on the time evolution of the off-diagonal element of qubit-1, which is written as
\begin{equation}
\rho_{1,\downarrow\uparrow}^\mathrm{I}(t)=\rho_{1,\downarrow\uparrow}(0)\left(\rho_{2,\uparrow\uparrow}(0)\phi^{(\mathrm{X})}(t)+\rho_{2,\downarrow\downarrow}(0)\phi^{(\mathrm{X})*}(t)\right),
\end{equation}
where $\rho_{1,\downarrow\uparrow}(0)$, $\rho_{2,\uparrow\uparrow}(0)$, and $\rho_{2,\downarrow\downarrow}(0)$ are the initial conditions for the two qubits
and the dephasing factor $\phi^{(\mathrm{X})}(t)$ is written as
\begin{equation}
\phi^{(\mathrm{X})}(t)=e^{-i2\left(\theta_{1,2}(t)+\theta_{2,1}(t)\right)}\langle\prod_{\vec{k}}\widehat{D}_{\vec{k},+}^\dagger(t)\widehat{D}_{\vec{k},-}^\dagger(t)\rangle,
\end{equation}
where
\begin{eqnarray}
\theta_{j,j'}(t)&=&\sum_{\vec{k}}g_{j,\vec{k}}g_{j',\vec{k}}^*\left(\frac{\omega_{\vec{k}}t-\sin\omega_{\vec{k}}t}{\omega_{\vec{k}}^2}\right) \nonumber\\
&=&\int_0^\infty\frac{\mathcal{J}_{j,j'}(\omega)}{\omega^2}\left(\omega t-\sin\omega t\right)d\omega,
\end{eqnarray}
$\mathcal{J}_{j,j'}(\omega)=\sum_{\vec{k}}g_{j,\vec{k}}g_{j',\vec{k}}^*\delta(\omega-\omega_{\vec{k}})$ are the spectral density functions, and
\begin{eqnarray}
\widehat{D}_{\vec{k},+}(t)&=&\exp\left[\left(g_{1,\vec{k}}+g_{2,\vec{k}}\right)\alpha_{\vec{k}}(t)\hat{b}_{\vec{k}}^\dagger\right. \nonumber\\
&&\left.-\left(g_{1,\vec{k}}+g_{2,\vec{k}}\right)^*\alpha_{\vec{k}}^*(t)\hat{b}_{\vec{k}}\right]  \nonumber\\
\widehat{D}_{\vec{k},-}(t)&=&\exp\left[\left(-g_{1,\vec{k}}+g_{2,\vec{k}}\right)\alpha_{\vec{k}}(t)\hat{b}_{\vec{k}}^\dagger\right. \nonumber\\
&&\left.-\left(-g_{1,\vec{k}}+g_{2,\vec{k}}\right)^*\alpha_{\vec{k}}^*(t)\hat{b}_{\vec{k}}\right]
\end{eqnarray}
represent the displacement operators, respectively.

The coupling constants $g_{j,\vec{k}}$ of the two qubits to the boson environment are in general complex numbers. In order to reveal the nonclassical effects caused by their relative phase, we assume, for simplicity, that they have the same amplitude, but with a phase difference:
\begin{equation}
g_{2,\vec{k}}=g_{1,\vec{k}} \, \exp[i\varphi].
\end{equation}

For a thermalized environment at temperature $T$, the two prescriptions
$\exp\left(\alpha\hat{b}^\dagger-\alpha^*\hat{b}\right)\exp\left(\beta\hat{b}^\dagger-\beta^*\hat{b}\right)=\exp\left[\left(\alpha\beta^*-\alpha^*\beta\right)/2\right]
\exp\left[(\alpha+\beta)\hat{b}^\dagger-(\alpha+\beta)^*\hat{b}\right]$
and
$\langle\exp\left(\alpha\hat{b}^\dagger-\alpha^*\hat{b}\right)\rangle
=\exp\left[-\coth(\hbar\omega/2k_\mathrm{B}T)|\alpha|^2/2\right]$,
are helpful for calculating the desired result
\begin{equation}
\phi^{(\mathrm{X})}(t)=\exp\left[-i\vartheta_\varphi(t)-\Phi(t)\right],
\label{Eq:Extended_model_dephasing_factor}
\end{equation}
where
\begin{eqnarray} \label{eq_vartheta}
\vartheta_\varphi(t)&=&\cos\varphi\int_0^\infty\frac{4\mathcal{J}(\omega)}{\omega^2}(\omega t-\sin\omega t)d\omega \\
&&+\mathrm{sign}(t)\sin\varphi\int_0^\infty\frac{4\mathcal{J}(\omega)}{\omega^2}(1-\cos\omega t)d\omega, \nonumber
\end{eqnarray}
$\mathcal{J}(\omega)=\sum_{\vec{k}}|g_{j,\vec{k}}|^2\delta(\omega-\omega_{\vec{k}})$ is the spectral density function, and
\begin{equation}
\Phi(t)=\int_0^\infty\frac{4\mathcal{J}(\omega)}{\omega^2}\coth\left(\frac{\hbar\omega}{2k_\mathrm{B}T}\right)(1-\cos\omega t)d\omega
\end{equation}
is the same as the one in the conventional spin-boson model. In the second line of Eq.~(\ref{eq_vartheta}), we have manually inserted $\mathrm{sign}(t)$. While this does not affect the pure dephasing dynamics for
$t\geq0$, it ensures that the condition $\phi^{(\mathrm{X})}(-t)=\phi^{(\mathrm{X})*}(t)$ is satisfied and one always obtains a real distribution
$\wp^{(\mathrm{X})}(\omega)$.

The presence of $\vartheta_\varphi(t)$ in Eq.~(\ref{Eq:Extended_model_dephasing_factor}) will in general result in the
violation of positivity. Note that, similar to the conventional spin-boson model, individual member Hamiltonians in the Hamiltonian ensemble must be of the form
$\omega\hat{\sigma}_\mathrm{z}/2$, which allows us to follow the same line of argument.

\section{OHMIC SPECTRAL DENSITY}

To demonstrate the violation of positivity explicitly, we consider the Ohmic spectral density function
\begin{equation}
\mathcal{J}_\mathrm{o1}(\omega)=\omega\exp(-\omega/\omega_\mathrm{c}) ,
\end{equation}
and the zero temperature limit where $T\rightarrow0$. For simplicity, we also assume a degenerate system Hamiltonian.

In the case of conventional spin-boson model, the dephasing factor is
\begin{equation}
\phi_\mathrm{o1}(t)=\frac{1}{\left(1+\omega_\mathrm{c}^2t^2\right)^2},
\end{equation}
and the corresponding distribution is
\begin{equation}
\wp_\mathrm{o1}(\omega)=\frac{1}{4\omega_\mathrm{c}^2}(\omega_\mathrm{c}+|\omega|)\exp[-\frac{|\omega|}{\omega_\mathrm{c}}].
\end{equation}
This is obviously a legitimate probability distribution without negative values. The results are shown in Fig.~2(a) of the main article. Consequently, the
Hamiltonian ensemble $\left\{\left(\omega\hat{\sigma}_\mathrm{z}/2,\wp_\mathrm{o1}(\omega)\right)\right\}$ resembles the same pure dephasing dynamics of the
conventional spin-boson model characterized by $\phi_\mathrm{o1}(t)$. As expected, $\wp_\mathrm{o1}(\omega)$ is, due to the degeneracy of the system Hamiltonian,
centered at $\omega=0$, and broadens with increasing $\omega_\mathrm{c}$.

Whereas, for the extended model, the dephasing factor reads
\begin{equation}
\phi_\mathrm{o1}^{(\mathrm{X})}(t)=\frac{\exp\left[-i4\cos\varphi\left(\omega_\mathrm{c}t-\arctan(\omega_\mathrm{c}t)\right)\right]}
{\left(1+\omega_\mathrm{c}^2t^2\right)^{2(1+i\mathrm{sign}(t)\sin\varphi)}}.
\end{equation}
Since the condition $\phi_\mathrm{o1}^{(\mathrm{X})}(-t)=\phi_\mathrm{o1}^{(\mathrm{X})*}(t)$ is fulfilled, the corresponding distribution
$\wp_\mathrm{o1}^{(\mathrm{X})}(t)$ is real. However, the positivity of $\wp_\mathrm{o1}^{(\mathrm{X})}(t)$ is in general lost due to the presence of the
nontrivial phase angle $\vartheta_\varphi(t)$. The results are show in Fig.~2(b) and (c) of the main article.

\end{document}